\documentclass[a4paper]{PoS}

\usepackage{textcomp}
\usepackage{amsfonts} 
\usepackage{amssymb} 
\usepackage{amsmath} 
\usepackage{slashed}
\usepackage{graphicx}
\usepackage{pifont}
\usepackage{multirow,lscape}
\usepackage{array}
\usepackage{longtable}
\usepackage[FIGBOTCAP,TABBOTCAP,tight]{subfigure}
\usepackage{verbatim}
\usepackage{bm}
\usepackage{comment}
\usepackage{color}
\usepackage{url} 

\graphicspath{{./figs/}}

\providecommand{\wbar}[1]{\overline#1}
\providecommand{\abs}[1]{\lvert#1\rvert}

\providecommand{\mate}[3]{\langle#1\lvert#2\rvert#3\rangle}

\renewcommand{\Re}{\mathrm{Re}}
\renewcommand{\Im}{\mathrm{Im}}

\newcolumntype{C}[1]{>{\centering\arraybackslash}p{#1}}

\newcommand{\BK}{\hat{B}_{K}}
\newcommand{\Vcb}{V_{cb}}

\newcommand{\eps}{\varepsilon}

\newcommand{\epsK}{\varepsilon_{K}}

\title{Determination of $\varepsilon_K$ using lattice QCD inputs}

\ShortTitle{Determination of $\varepsilon_K$}

\author{ Jon A.~Bailey, Yong-Chull Jang, \speaker{Weonjong Lee}, and
  Sungwoo Park\\
  Lattice Gauge Theory Research Center, CTP, and FPRD, \\
  Department of Physics and Astronomy, \\
  Seoul National University,
  Seoul 08826, South Korea\\
  E-mail: \email{wlee@snu.ac.kr} }

\author{SWME Collaboration}

\abstract{ We present results for the indirect CP violation parameter
  $\epsK$ determined directly from the standard model using lattice
  QCD to fix the inputs $\BK$, $\xi_0$, $|V_{us}|$, and $|V_{cb}|$.
  We use the FLAG and SWME results for $\BK$.  We use the RBC-UKQCD
  result for $\xi_0$ determined using the experimental value of
  $\eps'/\eps$ and the lattice result of $\Im A_2$. To set the
  Wolfenstein parameter $\lambda$, we use $|V_{us}|$, which is
  determined from $K_{\ell3}$ and $K_{\mu2}$ decays combined with
  lattice evaluations of the $K \to \pi \ell \nu$ vector form factor
  and $f_K$. To set the Wolfenstein parameter $A$, we use the
  FNAL/MILC results for $|V_{cb}|$, which are determined from the
  exclusive decay $\bar{B} \to D^* \ell \bar{\nu}$ and the axial form
  factor at zero recoil. We also use the inclusive $|V_{cb}|$ obtained
  using the heavy quark expansion based on QCD sum rules and the
  OPE. We compare the results with those for exclusive $|V_{cb}|$. We
  find that the standard model prediction of $\epsK$ with exclusive
  $|V_{cb}|$ (lattice QCD results) is lower than the experimental value
  by 3.4$\sigma$. However, we observe no tension in $\epsK$ determined
  from inclusive $|V_{cb}|$. }

\FullConference{The 33rd International Symposium on Lattice Field Theory\\
                 14 -18 July  2015\\
                 Kobe International Conference Center, Kobe, Japan}

\begin{document}

\section{Introduction}
\label{sec:intr}

Indirect CP violation in neutral kaons is parametrized by $\epsK$
\begin{equation}
  \label{eq:epsK_def}
  \epsK 
  \equiv \frac{\mathcal{A}(K_L \to \pi\pi(I=0))} 
              {\mathcal{A}(K_S \to \pi\pi(I=0))} \,.
\end{equation}
Here, $K_L$ and $K_S$ are the neutral kaon states in nature.
We can also calculate $\epsK$ directly from the standard model (SM)
using tools in lattice QCD.
Hence, we can test the SM through CP violation by comparing the
experimental and theoretical value of $\epsK$.

In order to calculate $\epsK$ directly from the SM, we use input
parameters obtained from lattice QCD and experiments.
In particular, we use lattice QCD inputs for $\BK$, $\abs{V_{cb}}$,
$\abs{V_{us}}$, and $\xi_0$ in this paper.
In addition, in order to avoid unwanted correlation through $\epsK$
between the Wolfenstein parameters of the CKM matrix and the inputs,
we adopt the angle-only fit (AOF) from the UTfit collaboration \cite{
  Bevan2013:npps241.89} to determine the apex
$(\bar{\rho},\bar{\eta})$ of the unitarity triangle.

\section{Master formula for $\epsK$}
\label{sec:master-f}

In the SM, the master formula for $\epsK$ is
\begin{align}
  \label{eq:epsK_SM_0}
  \epsK
  =& e^{i\theta} \sqrt{2}\sin{\theta} 
  \Big( C_{\eps} \; X_\text{SD} \; \hat{B}_{K} 
  + \frac{ \xi_{0} }{ \sqrt{2} } + \xi_\text{LD} \Big)
   + \mathcal{O}(\omega \eps^\prime)
   + \mathcal{O}(\xi_0 \Gamma_2/\Gamma_1) \,,
\end{align}
where
\begin{align}
  C_{\eps} 
  &= \frac{ G_{F}^{2} F_K^{2} m_{K^{0}} M_{W}^{2} }
  { 6\sqrt{2} \; \pi^{2} \; \Delta M_{K} } \,,
  \quad
  \xi_\text{LD} =  \frac{m^\prime_\text{LD}}{\sqrt{2} \; \Delta M_K} \,,
  \quad
  m^\prime_\text{LD}
  = -\Im \left[ \mathcal{P} \; \sum_{C} 
    \frac{\mate{\wbar{K}^0}{H_\text{w}}{C} \mate{C}{H_\text{w}}{K^0}}
         {m_{K^0}-E_{C}}
         \right]
  \label{eq:mLD}
\end{align}
Here, $X_\text{SD}$ is the short distance contribution from the box
diagrams:
\begin{align}
  X_\text{SD} &= \Im\lambda_t \Big[ \Re\lambda_c \eta_{cc} S_0(x_c)
    -\Re\lambda_t \eta_{tt} S_0(x_t) - (\Re\lambda_c -
    \Re\lambda_t) \eta_{ct} S_0(x_c,x_t) \Big] \,,
\end{align}
where $S_0$ are the Inami-Lim functions given in Ref.~\cite{
  Inami1980:ProgTheorPhys.65.297}, $\lambda_{i} \equiv V_{is}^{\ast}
V_{id} $, and $x_i = m_i^2/M_W^2$ with $m_i = m_i(m_i)$ defined as the
scale invariant $\overline{\text{MS}}$ quark mass \cite{
  Chetyrkin2000:CompPhysComm.133.43}.
The $\xi_0$ term represents the long distance effect from the
absorptive part of the effective Hamiltonian: $\xi_0 = \Im A_0/\Re
A_0$.
The $\xi_\text{LD}$ term represents the long distance effect from
the dispersive part of the effective Hamiltonian.
Details of how to derive the master formula in
Eq.~\eqref{eq:epsK_SM_0} directly from the standard model using
Wigner-Weisskopf perturbation theory are given in Ref.~\cite{
  Bailey:2015tba}.

\section{Input parameters}
\label{sec:in-para}

%
%
\begin{table}[t!]
  \renewcommand{\subfigcapskip}{0.55em}
  \subtable[Wolfenstein parameters]{
    \resizebox{0.63\textwidth}{!}{
      \begin{tabular}{cccc}
        \hline\hline
        & CKMfitter & UTfit & AOF \cite{Bevan2013:npps241.89} \\ \hline
        $\lambda$
        & $0.22537(61)$
        /\cite{Agashe2014:ChinPhysC.38.090001}
        & $0.2255(6)$
        /\cite{Agashe2014:ChinPhysC.38.090001}
        & $0.2253(8)$
        /\cite{Agashe2014:ChinPhysC.38.090001}
        \\ \hline
        $\bar{\rho}$
        & $0.117(21)$
        /\cite{Agashe2014:ChinPhysC.38.090001}
        & $0.124(24)$
        /\cite{Agashe2014:ChinPhysC.38.090001}
        & $0.139(29)$
        /\cite{UTfit2014PostMoriondSM:web}
        \\ \hline
        $\bar{\eta}$
        & $0.353(13)$
        /\cite{Agashe2014:ChinPhysC.38.090001}
        & $0.354(15)$
        /\cite{Agashe2014:ChinPhysC.38.090001}
        & $0.337(16)$
        /\cite{UTfit2014PostMoriondSM:web}
        \\ \hline\hline
      \end{tabular} 
    } 
    \label{tab:in-wolf}
  } 
  \hfill
  \subtable[QCD corrections]{
    \resizebox{0.32\textwidth}{!}{
      \begin{tabular}{clc}
        \hline\hline
        Input & Value & Ref. \\ \hline
        $\eta_{cc}$ & $1.72(27)$
        & { \protect\cite{Bailey:2015tba} } \\ \hline
        $\eta_{tt}$ & $0.5765(65)$
        & { \protect\cite{Buras2008:PhysRevD.78.033005} } \\ \hline
        $\eta_{ct}$ & $0.496(47)$
        & { \protect\cite{Brod2010:prd.82.094026} }
        \\ \hline\hline
      \end{tabular}
      } 
    \label{tab:in-eta}
  } 
  \\
  \subtable[$\abs{V_{cb}}$ in units of $10^{-3}$]{
    \resizebox{0.55\textwidth}{!}{
      \begin{tabular}{ccc}
        \hline\hline
        Inclusive (Kin.) & Inclusive (1S) & Exclusive \\ \hline
        $42.21(78)$
        /\cite{Alberti2014:PhysRevLett.114.061802}
        & $41.96(45)(07)$
        /\cite{Bauer2004:PhysRevD.70.094017}
        & $39.04(49)(53)(19)$
        /\cite{Bailey2014:PhysRevD.89.114504}
        \\ \hline\hline
      \end{tabular}
    } 
    \label{tab:in-Vcb}
  } 
  \hfill
  \subtable[$\BK$]{
    \resizebox{0.43\textwidth}{!}{
      \begin{tabular}{cccc}
        \hline\hline
        & FLAG & SWME & \\ \hline
        & $0.7661(99)$
        /{ \protect\cite{Aoki2013:hep-lat.1310.8555} }
        & $0.7379(47)(365)$
        /{ \protect\cite{Bae2014:prd.89.074504} }
        &
        \\ \hline\hline
      \end{tabular}
    } 
    \label{tab:in-BK}
  } 
  \\
  \subtable[Long distance effects]{
    \resizebox{0.45\textwidth}{!}{
      \begin{tabular}{clc}
        \hline\hline
        Input & Value & Ref. \\ \hline
        $\xi_0$ & $-1.63(19)(20) \times 10^{-4}$ 
        & { \protect\cite{Blum2011:PhysRevLett.108.141601} } \\ \hline
        $\xi_\text{LD}$ & $(0 \pm 1.6)\,\%$ 
        & { \protect\cite{Christ2012:PhysRevD.88.014508} } 
        \\ \hline\hline
      \end{tabular}
    } 
    \label{tab:in-LD}
  } 
  \hfill
  \subtable[Other input parameters]{
    \resizebox{0.48\textwidth}{!}{
      \begin{tabular}{clc}
        \hline\hline
        Input & Value & Ref. \\ \hline
        $G_{F}$
        & $1.1663787(6) \times 10^{-5}$ GeV$^{-2}$
        &\cite{Agashe2014:ChinPhysC.38.090001} \\ \hline
        $M_{W}$
        & $80.385(15)$ GeV
        &\cite{Agashe2014:ChinPhysC.38.090001} \\ \hline
        $m_{c}(m_{c})$
        & $1.275(25)$ GeV
        &\cite{Agashe2014:ChinPhysC.38.090001} \\ \hline
        $m_{t}(m_{t})$
        & $163.3(2.7)$ GeV
        &\cite{Alekhin2012:plb.716.214} \\ \hline
        $\theta$
        & $43.52(5)^{\circ}$
        &\cite{Agashe2014:ChinPhysC.38.090001} \\ \hline
        $m_{K^{0}}$
        & $497.614(24)$ MeV
        &\cite{Agashe2014:ChinPhysC.38.090001} \\ \hline
        $\Delta M_{K}$
        & $3.484(6) \times 10^{-12}$ MeV
        &\cite{Agashe2014:ChinPhysC.38.090001} \\ \hline
        $F_K$
        & $156.2(7)$ MeV
        &\cite{Agashe2014:ChinPhysC.38.090001}
        \\ \hline\hline
      \end{tabular}
    } 
    \label{tab:in-extra}
  } 
  \caption{ Input parameters }
  \label{tab:input}
\end{table}
%

The CKMfitter and UTfit groups provide the Wolfenstein  parameters
$\lambda$, $\bar{\rho}$, $\bar{\eta}$ and $A$ from the global
unitarity triangle (UT) fit, which are summarized in Table
\ref{tab:input}\,\subref{tab:in-wolf}.
Here, the parameters $\epsK$, $\BK$, and $\abs{V_{cb}}$ are inputs to the
global UT fit.
Hence, the Wolfenstein parameters extracted from the global UT fit
contain unwanted dependence on $\epsK$.
In order to avoid this unwanted correlation and to determine $\epsK$
self-consistently, we take another input set from the angle-only fit
(AOF) in Ref.~\cite{Bevan2013:npps241.89}.
The AOF does not use $\epsK$, $\BK$, or $\abs{V_{cb}}$ as input to determine
the UT apex $(\bar{\rho},\bar{\eta})$.
We take $\lambda$ independently from $\abs{V_{us}}$, which has been extracted
from the $K_{\ell 3}$ and $K_{\mu2}$ decays with lattice QCD inputs
\cite{Agashe2014:ChinPhysC.38.090001}.

The input values for $\abs{V_{cb}}$ are summarized in Table
\ref{tab:input}\,\subref{tab:in-Vcb}.
The inclusive determination takes into account the inclusive decay
modes: $B \to X_c\ell\nu$ (essential) and $B \to X_s \gamma$
(optional).
Moments of lepton energy, hadron masses, and photon energy (optional)
are measured from the relevant decays.
These moments are fit to the theoretical formula which is a heavy
quark expansion obtained with the aid of the operator product
expansion (OPE) \cite{ Agashe2014:ChinPhysC.38.090001,
  Alberti2014:PhysRevLett.114.061802}.
Here, we use the most updated value, given in Ref.~\cite{
  Alberti2014:PhysRevLett.114.061802}.
%

For the exclusive $\abs{V_{cb}}$, we use the most precise value from
the FNAL/MILC lattice calculation of the form factor $\mathcal{F}(w)$
of the semileptonic decay $\bar{B}\to D^*\ell\bar{\nu}$ at zero recoil
($w=1$) \cite{ Bailey2014:PhysRevD.89.114504}.
They combined their lattice result with the HFAG average \cite{
  Amhis2012:HFAG} of $\mathcal{F}(1) \abs{\bar{\eta}_\text{EM}}
\abs{V_{cb}}$ to extract $\abs{V_{cb}}$.

There have been a number of lattice QCD calculations of $\BK$ with
$N_f=2+1$ \cite{ Bae2012:PhysRevLett.109.041601,
  Durr2011:PhysLettB.705.477, Aubin2010:PhysRevD.81.014507,
  Arthur:2012yc, Blum:2014tka}.
Here, we use the FLAG average in Ref.~\cite{
  Aoki2013:hep-lat.1310.8555} and the SWME result in Ref.~\cite{
  Bae2014:prd.89.074504}, which deviates most from the FLAG average.
They are summarized in Table \ref{tab:input}\,\subref{tab:in-BK}.

The RBC/UKQCD collaboration provides lattice results for $\Im A_2$ and
$\xi_0$ in Ref.~\cite{ Blum2011:PhysRevLett.108.141601, Blum:2015ywa}.
The long distance effect $\xi_0$ is given in Table
\ref{tab:input}\,\subref{tab:in-LD}.
In the master formula in Eq.~\eqref{eq:epsK_SM_0}, $\xi_\text{LD}$
represents the long distance effect of $\approx 2\%$ which comes from
the dispersive part of the effective Hamiltonian.
The precise evaluation of $\xi_\text{LD}$ from lattice QCD is not
available yet.
Hence, we do not include this effect in the central value of $\epsK$,
but we take it as a systematic error with the value given in Table
\ref{tab:in-LD}.
The correction terms $\mathcal{O}(\omega \eps^\prime)$ and
$\mathcal{O}(\xi_0 \Gamma_2/\Gamma_1)$ are of order $10^{-7}$, and we
neglect them in this analysis.
A rough estimate of $\xi_\text{LD}$ is available from Ref.~\cite{
  Christ2012:PhysRevD.88.014508}.

The $\eta_{ij}$ parameters in Table
\ref{tab:input}\,\subref{tab:in-eta} represent the QCD corrections to
the coefficients of Inami-Lim functions.
The factor $\eta_{tt}$ is given at NLO, whereas $\eta_{cc}$ and
$\eta_{ct}$ are known up to NNLO.
Refer to Ref.~\cite{ Bailey:2015tba} for more details.
%
The rest of the input parameters are given in Table
\ref{tab:input}\,\subref{tab:in-extra}.

\section{Results}
\label{sec:res}

Let us define $\epsK^\text{SM}$ as the theoretical evaluation of
$\abs{\epsK}$ using the master formula of Eq.~\eqref{eq:epsK_SM_0}.
We define $\epsK^\text{Exp}$ as the experimental value of
$\abs{\epsK}$: $\epsK^\text{Exp} = (2.228 \pm 0.011) \times 10^{-3}$
\cite{ Agashe2014:ChinPhysC.38.090001}.
Let us define $\Delta \epsK$ as the difference between
$\epsK^\text{Exp}$ and $\epsK^\text{SM}$: $\Delta\epsK \equiv
\epsK^\text{Exp} - \epsK^\text{SM}$.
Here, we assume that the theoretical phase $\theta$ is equal to the
experimental phase $\phi_\varepsilon$, although it is not fully
confirmed in lattice QCD yet.

\begin{figure}[t!]
  \subfigure[Exclusive $\abs{V_{cb}}$]{
    \includegraphics[width=0.48\textwidth]{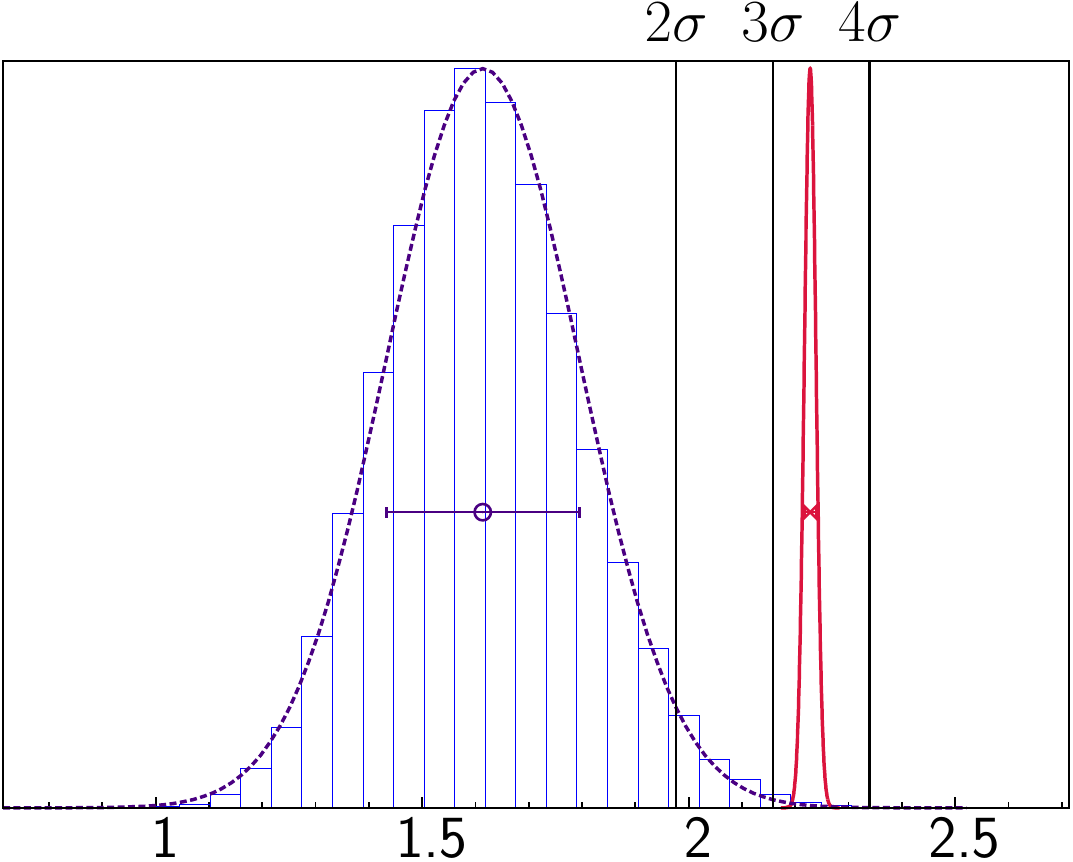}
    \label{sfig:excl.Vcb}
  }
  \hfill
  \subfigure[Inclusive $\abs{V_{cb}}$]{
    \centering
    \includegraphics[width=0.48\textwidth]{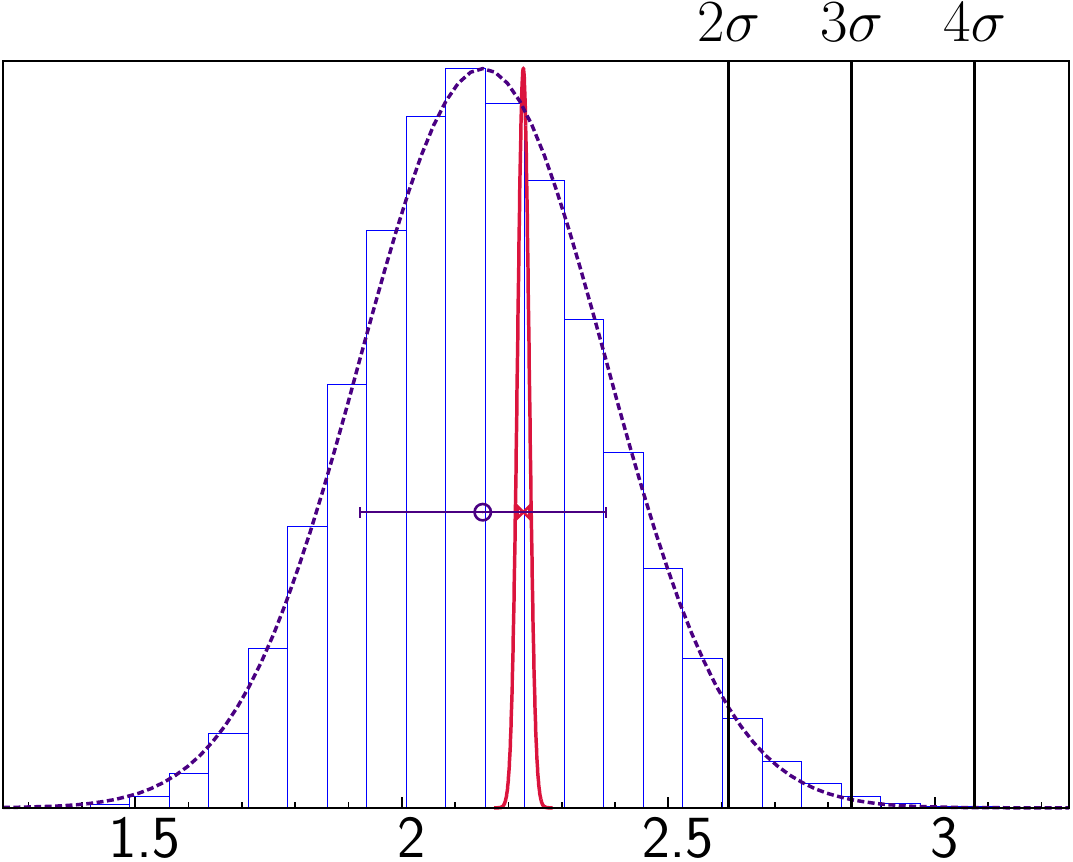}
    \label{sfig:incl.Vcb}
  }
  \caption{ Gaussian probability distributions for $\epsK^\text{SM}$
    (blue dotted line) and $\epsK^\text{Exp}$ (red solid line).  Here,
    the results are obtained using the FLAG $\BK$ and the AOF
    parameters for the CKM matrix elements. }
  \label{fig:p-dist}
\end{figure}

In Table \ref{tab:epsK+DepsK}\,\subref{tab:epsK}, we present results 
for $\epsK^\text{SM}$.
They are obtained using the FLAG average for $\BK$ \cite{
  Aoki2013:hep-lat.1310.8555}, inclusive $\abs{V_{cb}}$ from
Ref.~\cite{ Alberti2014:PhysRevLett.114.061802}, and exclusive
$\abs{V_{cb}}$ from Ref.~\cite{ Bailey2014:PhysRevD.89.114504}.
The corresponding probability distributions for $\epsK^\text{SM}$ are
presented in Fig.~\ref{fig:p-dist}.

%
\begin{table}[t!]
  \renewcommand{\subfigcapskip}{0.55em}
  \subtable[$\epsK^\text{SM}$]{
    \resizebox{0.48\textwidth}{!}{
      \begin{tabular}{ccc}
        \hline\hline
        Input Method & Inclusive $\Vcb$ & Exclusive $\Vcb$
        \\ \hline
        CKMfitter
        & $2.31(23)$ 
        & $1.73(18)$ 
        \\ \hline
        UTfit
        & $2.30(24)$ 
        & $1.73(19)$ 
        \\ \hline
        AOF
        & $2.15(23)$ 
        & $1.61(18)$ 
        \\ \hline\hline
      \end{tabular}
    } 
    \label{tab:epsK}
  } 
  \hfill
  \subtable[$\Delta\epsK$]{
    \resizebox{0.48\textwidth}{!}{
    \begin{tabular}{ccc}
      \hline\hline
      Input Method &
      Inclusive $\Vcb$ &
      Exclusive $\Vcb$
      \\ \hline
      CKMfitter
      & $-0.34\sigma$ 
      & $2.7\sigma$ 
      \\ \hline
      UTfit
      & $-0.31\sigma$ 
      & $2.7\sigma$ 
      \\ \hline
      AOF
      & $0.33\sigma$ 
      & $3.4\sigma$ 
      \\ \hline\hline
    \end{tabular}
    } 
    \label{tab:DepsK}
  } 
  \caption{ \protect\subref{tab:epsK} $\epsK^\text{SM}$ in units of $10^{-3}$,
    and \protect\subref{tab:DepsK} $\Delta\epsK$ in units of $\sigma$. The
    $\sigma$ is obtained by combining errors of $\epsK^\text{SM}$ and
    $\epsK^\text{Exp}$ in quadrature. }
  \label{tab:epsK+DepsK}
\end{table}

In Table \ref{tab:epsK+DepsK}\,\subref{tab:DepsK}, we present results
for $\Delta\epsK$ for both inclusive and exclusive $\abs{V_{cb}}$.
From Table \ref{tab:epsK+DepsK}, we observe no tension in
$\Delta\epsK$ in the inclusive decay channels for $\abs{V_{cb}}$,
which are obtained using QCD sum rules and the heavy quark expansion.
However, from Table \ref{tab:epsK+DepsK}, we find that there exists a
$3.4\sigma$ tension between $\epsK^\text{Exp}$ and $\epsK^\text{SM}$
obtained using the exclusive $\abs{V_{cb}}$, which is determined using
lattice QCD tools.
In other words, $\epsK^\text{SM}$ with exclusive $\abs{V_{cb}}$ and
the most reliable input method (AOF) is only 72\% of
$\epsK^\text{Exp}$.
The largest contribution that we neglect in our estimate of
$\epsK^\text{SM}$ is much less than 2\%.
Hence, the neglected contributions cannot explain the gap
$\Delta\epsK$ of 28\% with exclusive $\abs{V_{cb}}$.
%
%

%

\begin{figure}[b!]
  \centering
  \includegraphics[width=0.5\columnwidth]{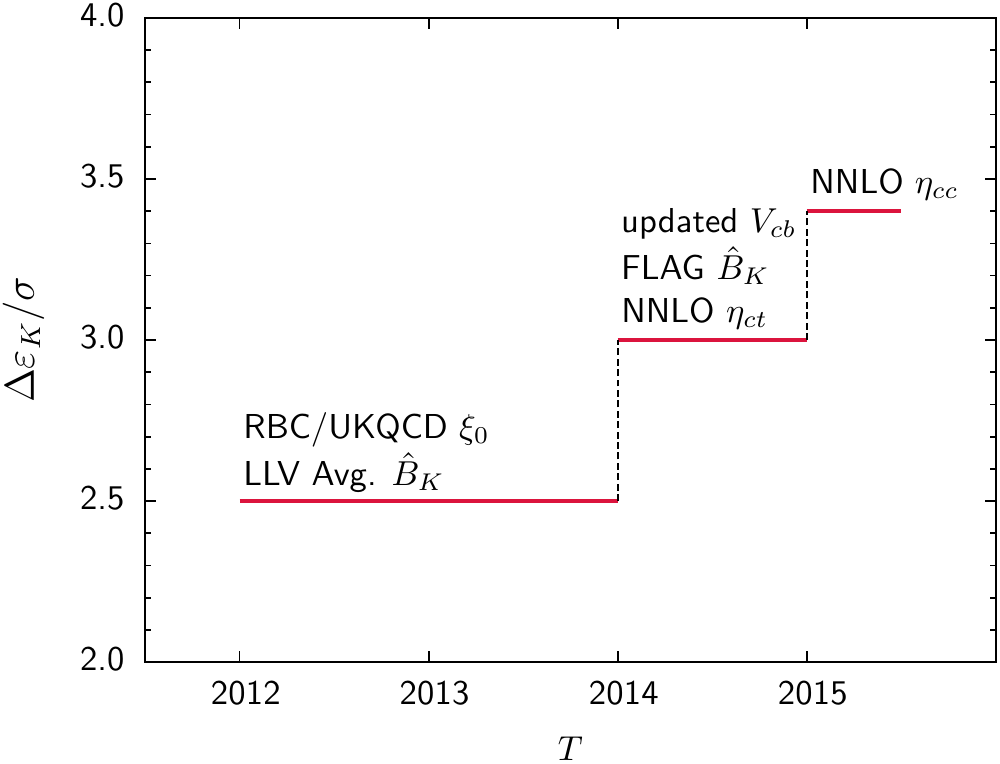}
  \caption{Recent history of $\Delta \epsK$ along with the theoretical
    progress.}
  \label{fig:history}
\end{figure}

In Fig.~\ref{fig:history}, we present the chronological evolution of
$\Delta\epsK/\sigma$ as the progress in lattice and perturbative QCD
goes on.
In 2012, RBC/UKQCD reported $\xi_0$ in Ref.~\cite{
  Blum2011:PhysRevLett.108.141601}, and the lattice average for $\BK$
by LLV became available in Ref.~\cite{ Laiho2009:PhysRevD.81.034503}.
Based on these works, SWME reported $\Delta \epsK = 2.5\sigma$ in
Ref.~\cite{ Jang2012:PoS.LAT2012.269} in 2012.
The FLAG average for $\BK$ became available in Ref.~\cite{
  Aoki2013:hep-lat.1310.8555} in 2013.
In 2014, FNAL/MILC reported an updated $\abs{V_{cb}}$ in the exclusive
decay channel, and the NNLO value of $\eta_{ct}$ in Ref.~\cite{
  Brod2010:prd.82.094026} became known to us.
In 2014, SWME reported the updated $\Delta \epsK = 3.0\sigma$ in Ref.~\cite{
  Bailey:2014qda}.
In 2015, a remaining issue on the NNLO calculation of $\eta_{cc}$ was
addressed in Refs.~\cite{ Bailey:2015tba,
  Brod2011:PhysRevLett.108.121801, Buras2013:EurPhysJC.73.2560}.
In 2015, SWME reported the updated $\Delta \epsK = 3.4\sigma$ in
Ref.~\cite{ Bailey:2015tba}.

\section{Conclusion}
\label{sec:close}

Here, we find that there is a substantial $3.4\sigma$ tension in
$\epsK$ between experiment and the SM with lattice QCD inputs.
For the SM estimate of $\epsK$, we use the AOF parameters and lattice
QCD inputs for exclusive $\abs{V_{cb}}$, $\BK$, $\abs{V_{us}}$ and
$\xi_0$.
Since the AOF Wolfenstein parameters do not have unwanted correlation
with the lattice inputs via $\epsK$, the AOF method is relevant to the
data analysis in this paper.
We also find that the tension disappears for the inclusive
$\abs{V_{cb}}$, which is determined using QCD sum rules and the heavy
quark expansion.

In Table \ref{tab:epsK-budget}, we present the error budget for
$\epsK^\text{SM}$. 
In the second column of the tables, we show the fractional
contribution of each input parameter to the total error of
$\epsK^\text{SM}$.
From this error budget, we find that $\abs{V_{cb}}$ dominates the
error in $\epsK^\text{SM}$.
Therefore, it is essential to reduce the error of $\abs{V_{cb}}$ down
to the sub-percent level.
For this purpose, we plan to extract $\abs{V_{cb}}$ from the exclusive
channel using the Oktay-Kronfeld (OK) action \cite{
  Oktay2008:PhysRevD.78.014504} for heavy quarks to calculate the form
factors for $\bar{B} \to D^{(*)} \ell \bar{\nu}$ decays.
The first stage ground work for this goal is underway and preliminary
results are reported in Ref.~\cite{ Jang:LAT2014, Jang:LAT2015}.

Several lattice QCD inputs are obtained in the isospin limit, $m_u =
m_d$.
In particular, the isospin breaking effect from $\eps'/\eps$ in
$\xi_0$ could be substantial \cite{ Cirigliano:2003nn,
  Cirigliano:2003gt, Gardner:1999wb}.
The isospin breaking effects on $\xi_0$ and other input parameters are
of order 1\% in $\epsK$.
Here we neglect them, but will incorporate them into the evaluation of
$\eps_K$ in the future.
%

\begin{table}[tb!]
  \renewcommand{\subfigcapskip}{0.55em}
  \subtable[First]{
    \resizebox{0.40\textwidth}{!}{
      \begin{tabular}{ccc}
        \hline\hline
        source       & error (\%) & memo \\
        \hline
        $V_{cb}$     & 39.3        & FNAL/MILC \\
        $\bar{\eta}$ & 20.4        & AOF \\
        $\eta_{ct}$  & 16.9        & $c-t$ Box \\
        $\eta_{cc}$  &  7.1        & $c-c$ Box \\
        $\bar{\rho}$ &  5.4        & AOF \\
        $m_t$        &  2.4        & \\
        \hline\hline
      \end{tabular}
    } 
    \label{tab:err-1}
  } 
  \hfill
  \subtable[Second]{
    \resizebox{0.40\textwidth}{!}{
      \begin{tabular}{ccc}
        \hline\hline
        source       & error (\%) & memo \\
        \hline
        $\xi_0$      &  2.2        & RBC/UKQCD\\
        $\xi_\text{LD}$      &  2.0        & RBC/UKQCD\\
        $\hat{B}_K$  &  1.5        & FLAG \\
        $m_c$        &  1.0        & \\
        $\vdots$     & $\vdots$    & \\
        $\vdots$     & $\vdots$    & \\
        \hline\hline
      \end{tabular}
    } 
    \label{tab:err-2}
  } 
  \caption{ Error budget for $\epsK^\text{SM}$ obtained using the AOF
    method, the exclusive $\Vcb$, and the FLAG $\BK$. Here, the values
    are fractional contributions to the total error obtained using the
    formula given in Ref.~\cite{ Bailey:2015tba}.}
  \label{tab:epsK-budget}
\end{table}
%
%
%


\acknowledgments
Y.C.J. thanks A.~Soni for helpful discussion on the unitarity triangle
analysis.
We thank J.~Brod and A.~J. Buras for a helpful discussion on
$\eta_{cc}$.
The research of W.~Lee is supported by the Creative Research
Initiatives Program (No.~2015001776) of the NRF grant funded by the
Korean government (MEST).
J.A.B. is supported by the Basic Science Research Program of the
National Research Foundation of Korea (NRF) funded by the Ministry of
Education (No.~2015024974).
W.~Lee would like to acknowledge the support from the KISTI
supercomputing center through the strategic support program for the
supercomputing application research (No.~KSC-2014-G3-002).
Computations were carried out on the DAVID GPU clusters at Seoul
National University.

\bibliography{refs}



\end{document}